\documentclass{article}
\usepackage{float}
\usepackage{authblk}
\usepackage{algorithm}
\usepackage{algorithmicx}  
\usepackage{algpseudocode}  
\usepackage{setspace}
\usepackage{bm}

\usepackage{url}
\usepackage{amsmath}
\usepackage{stfloats}
\usepackage{color}
\usepackage{mathrsfs}
\usepackage{dcolumn}
\usepackage{subfigure}  
\usepackage{multirow}
\usepackage{graphics}
\usepackage{graphicx}
\usepackage{geometry}
\usepackage{amsfonts} 
\usepackage{amssymb}
\usepackage{amsthm}
\usepackage{booktabs}
\usepackage{longtable}

\definecolor{gry}{rgb}{0.6,0.6,0.6} 
\definecolor{mygreen}{rgb}{0,0.7,0} 
\usepackage[colorlinks,linkcolor=mygreen,anchorcolor=blue,citecolor=blue]{hyperref}
\title{\textbf{Probabilistic Reach-Avoid Reachability in Nondeterministic Systems with Time-Varying Targets and Obstacles}}

\author[a,b]{Wei Liao}
\author[a,b]{Taotao Liang}
\author[a,b]{Xiaohui Wei \thanks{Corresponding author: wei\_xiaohui@nuaa.edu.cn}}
\author[a,b]{Qiaozhi Yin}

\affil[a]{\footnotesize{ Key laboratory of Fundamental Science for National Defense-Advanced Design Technology of Flight Vehicle, 
Nanjing University of Aeronautics and Astronautics, Nanjing, Jiangsu, China }}
\affil[b]{\footnotesize{State Key Laboratory of Mechanics and Control of Mechanical Structures, 
Nanjing University of Aeronautics and Astronautics, Nanjing, Jiangsu, China}}

\date{}

\newtheorem{thm}{\textbf{Theorem}}

\newtheorem{definition}{\textbf{Definition}}
\begin{document}
\maketitle
\begin{abstract}
    The probabilistic reachability problems of nondeterministic systems are studied.
    Based on the existing studies, the definition of probabilistic reachable sets is 
    generalized by taking into account time-varying target set and obstacle.
    A numerical method is proposed to compute probabilistic reachable sets. 
    First, a scalar function in the state space is constructed by backward recursion and grid interpolation, 
    and then the probability reachable set is represented as a nonzero level set of this scalar function.
    In addition, based on the constructed scalar function, the optimal control policy can be designed.
    At the end of this paper, some examples are taken
    to illustrate the validity and accuracy of the proposed method.
\end{abstract}
\quad \\
\textbf{Keywards:}  Nondeterministic system, Probabilistic reachable set, Optimal control, Dynamic programming

\section{Introduction}
Reachability is an important property that describes the behavior of control systems.  
In a classic reachability problem, one
specifies a target set in the state space and then aims to find a set of initial states of the trajectories that can reach the target set
within a given time horizon \cite{a1,a2,a2.5}. Such a set is referred to as the reachable set. 
Such problems are of great interest in engineering, for example, in those involving feasibility \cite{a3} or safety \cite{a4}, 
where feasible or safe system states sometimes refer to those states that can reach the target set in a given duration.

So far, most researches on reachability are based on deterministic and exact modeling of the system, 
and they discuss a "yes" or "no" problem, 
i.e. whether it is possible to reach the target set in a given time starting from a certain set.
Various methods have been proposed to deal with such problems, including the ellipsoidal method \cite{a2.5,a5,a6}, 
the polyhedral method \cite{i1,a7}, and the level set method \cite{a1,a2,i2}.
The level set method requires less form for dynamical systems and can be used to solve nonlinear problems. 
Moreover, several mature toolboxes have been developed based on the level set method \cite{a8,a9,a10}, therefore, 
the level set method has become the most widely used method and is 
applied in a large number of fields such as flight control systems \cite{a13,a14}, 
ground traffic management systems \cite{i3,a12} and air traffic management systems \cite{a11,a15}.
In recent years, time-varying target sets and obstacles have been considered in several studies on the reachability of 
deterministic systems, allowing reachability analysis to be used for more complex engineering problems.
For instance, a time-varying obstacle can be regarded as a time-varying unsafe region outside of which the evolution trajectory 
of the system should be kept \cite{i4,a15,a16}.

However, in some engineering problems, it is quite difficult to model the system accurately. 
In these problems, due to the uncertainty of the systems themselves and external disturbances, 
the state transitions of the systems are no longer deterministic, but take the form of probability distributions.
In such cases, the reachability problem can no longer be simply answered by "yes" or "no". 
Although some studies have discussed the reachability of nondeterministic systems and 
proposed the definition of probabilistic reachable set \cite{a17,a18,a19}, 
these studies assume that the target set is time-invariant and do not take into account obstacles.

Motivated by the previous works mentioned above, the following contributions are made in this paper:
\begin{itemize}
    \item [(1)] In this paper, we adopt a discrete time point of view to study reachability problems of nondeterministic systems, 
    in which we consider time-varying target sets and obstacles, and refine the definition of probabilistic reachable sets.
    \item [(2)] A method based on recursion and grid interpolation is proposed for computing probabilistic reachable sets. In this method, 
    the probabilistic reachable set is represented as a level set of a scalar function approximated by grid interpolation.
\end{itemize}

This paper unfolds as follows: Section 2 mathematically describes the problem. 
The principle of the method for computing probabilistic reachable sets is presented in Section 3. 
Section 4 introduces the implementation of the method. 
A numerical examples are given in Section 5 to illustrate the validity of the proposed method.
The results are summarized in Section 6.

\section{Problem Statement}
Consider the following discrete-time control system with uncertainty:
\begin{align}\label{eq1}
    s_{k+1}\sim \mathcal{F}\left(.|s_k,u_k\right)
\end{align}
where $s_k\in \mathbb{R}^n$ and $u_k$ are the system state ant control input at time $k$, respectively,
and $u_k$ is selected from a prescribed set $\mathcal{U}$. 
$\mathcal{F}\left(.|s_k,u_k\right)$ denotes a probability distribution which is related to $s_k$ and $u_k$,
and for any $s\in\mathbb{R}^n$, $\mathcal{F}\left(s|s_k,u_k\right)$ represents the probability density of 
this probability distribution at $s$.

\begin{definition}[Control policy]\label{def1}
    The sequence of control inputs $(u_0,u_1,...)$ can be determined by a control policy which can be represented by a mapping 
    $\mathcal{E}(.,.):\mathbb{R}^n\times \mathbb{N}\to \mathcal{U}$, and $u_k=\mathcal{E}(s_k,k)$.
\end{definition}

Let $\mathbb{Z}_{k_1}^{k_2}$ ($k_1$, $k_2$ are integers and $k_2\geq k_1$) denote the set of all integers between $k_1$ and $k_2$,
i.e., $\mathbb{Z}_{k_1}^{k_2}=\{k_1,k_1+1,...,k_2\}$ and let $\mathcal{D}$ denote the set of all probability distributions on $\mathbb{R}^n$.
Then, given the state $s_{k_1}$ at time $k_1$ and control policy $\mathcal{E}(.,.)$, the evolution of system 
(\ref{eq1}) in $\mathbb{Z}_{k_1}^{k_2}$ can be expressed as a probability distribution over time 
$\phi_{k_1}^{k_2}\left(.|.,s_{k_1},\mathcal{E}\right):\mathbb{Z}_{k_1}^{k_2}\to \mathcal{D}$, and 
$\phi_{k_1}^{k_2}\left(.|k,s_{k_1},\mathcal{E}\right)$ denotes the distribution of the system state at time $k$, 
while $\phi_{k_1}^{k_2}\left(s|k,s_{k_1},\mathcal{E}\right)$ denotes the probability density of the system state reaching state $s$ at time $k$.
Denote by $\mathcal{A}_k$ and $\mathcal{B}_k$ the target set and the obstacle at time $k$, respectively. Consequently,
given a time horizon $T\in\mathbb{N}$ and a control policy $\mathcal{E}(.,.)$, 
The evolution of system (\ref{eq1}) initializing from any state $s_0\in \mathbb{R}^n$ 
has a probability of reaching the target set at some time $k\in \mathbb{Z}_{0}^T$ and avoiding obstacles before 
reaching the target set, and this probability is expressed mathematically as:
\begin{align}\label{eq2}
    \mathrm{P}\left(s_0|{T},{\mathcal{E}}\right)=
    \mathrm{P}\left[ \left(\exists k\in\mathbb{Z}_0^T, 
            \phi_0^T(.|k,s_0,\mathcal{E} )\in \mathcal{A}_k \right) 
            \land 
            \left(\forall l\in \mathbb{Z}_0^k, \phi_0^T(.|l,s_0,\mathcal{E} )\notin \mathcal{B}_l \right) \right]
\end{align}
where "$\land$" is the logical operator "AND".
Then, given a positive real number $\gamma\in [0,1]$, the probabilistic reachable set can be defined:
\begin{definition}[$\gamma-$probabilistic reachable set]\label{def2}
    \begin{align}\label{eq3}
        \mathcal{R}_\gamma^{\mathcal{E}}(\mathcal{A},\mathcal{B},T)=\left\{ s_0\in\mathbb{R}^n|\mathrm{P}\left(s_0|{T},{\mathcal{E}}\right)\geq \gamma   \right\}
    \end{align}
\end{definition}
It is clear that the value of the expression of the probability in Eq. (\ref{eq2}) depends on the control policy adopted.
By finding the optimal control policy one can maximize this probability and the maximum probabilistic reachable set can be defined:
\begin{definition}[Maximum $\gamma-$probabilistic reachable set]\label{def3}
    \begin{align}\label{eq4}
        \mathcal{R}_\gamma^{\mathrm{max}}(\mathcal{A},\mathcal{B},T)=\left\{ s_0\in\mathbb{R}^n|\max_{\mathcal{E}} \mathrm{P}\left(s_0|{T},{\mathcal{E}}\right)\geq \gamma   \right\}
    \end{align}
\end{definition}
In the next section, we show that $\mathcal{R}_\gamma^{\mathcal{E}}(\mathcal{A},\mathcal{B},T)$ and 
$\mathcal{R}_\gamma^{\mathrm{max}}(\mathcal{A},\mathcal{B},T)$ can be computed by a backward recursive procedure.

\section{Method to compute probabilistic reachable sets}
Let $\mathbb{I}_{\mathcal{A}_k}(.):\mathbb{R}^n\to \{0,1\}$ and $\mathbb{O}_{\mathcal{B}_k}(.):\mathbb{R}^n\to \{0,1\}$
be the indicator functions of target set $\mathcal{A}_k$ and obstacle $\mathcal{B}_k$, respectively, i.e.:
\begin{align}\label{eq5}
  \mathbb{I}_{\mathcal{A}_k}(s)=\begin{cases}
    1,s\in \mathcal{A}_k\\
    0,s\notin \mathcal{A}_k
  \end{cases}\quad \quad 
  \mathbb{O}_{\mathcal{B}_k}(s)=\begin{cases}
    0,s\in \mathcal{B}_k\\
    1,s\notin \mathcal{B}_k
  \end{cases}
\end{align}
Then the following theorem can be proved. 

\begin{thm}\label{thm1}
    Fix a control policy $\mathcal{E}(.,.)$ and let the scalar function $V_k^{\mathcal{E}}(s_k)$
    denote the probability that the evolution of the system from state $s_k$ at time $k$ 
    reaches the target set at some time in $\mathbb{Z}_k^T$ and avoids the obstacle, i.e.:
    \begin{align}
        \begin{split}
        V_k^{\mathcal{E}}(s_{k})=  \mathrm{P} \left[ \left(\exists k_1\in \mathbb{Z}_k^T,\phi_k^T(.|k_1,s_k,\mathcal{E})\in \mathcal{A}_{k_1}\right) \land 
        \left(\forall k_2\in \mathbb{Z}_k^{k_1},\phi_k^{k_1}(.|k_2,s_k,\mathcal{E})\notin \mathcal{B}_{k_2}\right)  \right]
        \end{split}
    \end{align}
    If 
    \begin{align}
        \begin{split}
            V_{k-1}^{\mathcal{E}}(s_{k-1})=\min \left\{\mathbb{O}_{\mathcal{B}_{k-1}}(s_{k-1}), 
            \max\left\{\mathbb{I}_{\mathcal{A}_{k-1}}(s_{k-1}),
            \mathbb{E}\left[ V_k^{\mathcal{E}}\left( \mathcal{F}\left(.|s_{k-1},\mathcal{E}(s_{k-1},k-1) \right) \right)\right]  \right\} \right\}
        \end{split}
    \end{align}
    Then 
    \begin{align}\label{eq8}
        \begin{split}
        V_{k-1}^{\mathcal{E}}(s_{k-1})=  \mathrm{P} \left[ \left(\exists k_1\in \mathbb{Z}_{k-1}^T,
        \phi_{k-1}^T(.|k_1,s_{k-1},\mathcal{E})\in \mathcal{A}_{k_1}\right) \land 
        \left(\forall k_2\in \mathbb{Z}_{k-1}^{k_1},\phi_{k-1}^{k_1}(.|k_2,s_{k-1},\mathcal{E})\notin \mathcal{B}_{k_2}\right)  \right]
        \end{split}
    \end{align}
\end{thm}

\noindent
\textbf{Proof.} The event
\begin{align}
    \begin{split}
        \left(\exists k_1\in \mathbb{Z}_{k-1}^T,
        \phi_{k-1}^T(.|k_1,s_{k-1},\mathcal{E})\in \mathcal{A}_{k_1}\right) \land 
        \left(\forall k_2\in \mathbb{Z}_{k-1}^{k_1},\phi_{k-1}^{k_1}(.|k_2,s_{k-1},\mathcal{E})\notin \mathcal{B}_{k_2}\right) 
    \end{split}
\end{align}
and the event
\begin{align}
    \begin{split}
        &\left\{ \left( s_{k-1}\in \mathcal{A}_{k-1} \right)\lor 
        \left[ \left(\exists k_1\in \mathbb{Z}_{k}^T,\phi_k^T(.|k_1,s_{k-1},\mathcal{E})\in \mathcal{A}_{k_1}\right) \land 
        \left(\forall k_2\in \mathbb{Z}_k^{k_1},\phi_{k-1}^{k_1}(.|k_2,s_{k-1},\mathcal{E})\notin \mathcal{B}_{k_2}\right)  \right] \right\}  \\
        &\land \left( s_{k-1}\notin \mathcal{B}_{k-1} \right)
    \end{split}
\end{align}
are equivalent, where "$\lor$" is the logical operator "OR". 
For the sake of brevity, we let 
\begin{align}\label{eq11}
    \begin{split}
        &e_0^{\mathcal{E}}=\left(\exists k_1\in \mathbb{Z}_{k-1}^T,
        \phi_{k-1}^T(.|k_1,s_{k-1},\mathcal{E})\in \mathcal{A}_{k_1}\right) \land 
        \left(\forall k_2\in \mathbb{Z}_{k-1}^{k_1},\phi_{k-1}^{k_1}(.|k_2,s_{k-1},\mathcal{E})\notin \mathcal{B}_{k_2}\right)\\
        &e_1= s_{k-1}\in \mathcal{A}_{k-1}\\
        &e_2^{\mathcal{E}}= \left(\exists k_1\in \mathbb{Z}_{k}^T,\phi_{k-1}^T(.|k_1,s_{k-1},\mathcal{E})\in \mathcal{A}_{k_1}\right) \land 
        \left(\forall k_2\in \mathbb{Z}_k^{k_1},\phi_{k-1}^{k_1}(.|k_2,s_{k-1},\mathcal{E})\notin \mathcal{B}_{k_2}\right)\\
        &e_3=s_{k-1}\notin \mathcal{B}_{k-1}
    \end{split}
\end{align}
Therefore, $V_{k-1}^{\mathcal{E}}(s_{k-1})$ can be expanded as follows:
\begin{align}
    \begin{split}
        V_{k-1}^{\mathcal{E}}(s_{k-1})=\mathrm{P} \left(e_0^{\mathcal{E}}\right)
        = \left\{ 1- \left[ 1-\mathrm{P}\left( e_1 \right) \right] \times  
        \left[1-\mathrm{P}\left( e_2^{\mathcal{E}}  \right) \right] \right\}\times 
        \mathrm{P}\left( e_3 \right)
    \end{split}
\end{align}
According to the following equation, the state space can be divided into three parts:
\begin{align}
    \mathbb{R}^n = \mathcal{B}_{k-1} \cup \left( \mathcal{A}_{k-1}-\mathcal{B}_{k-1} \right) \cup \complement_{\mathbb{R}^n} \left( \mathcal{A}_{k-1}\cup\mathcal{B}_{k-1} \right)
\end{align}
where $\mathcal{A}_{k-1}-\mathcal{B}_{k-1}=\left\{s\in\mathbb{R}^n|s\in \mathcal{A}_{k-1} \land s\notin \mathcal{B}_{k-1} \right\}$.
These three parts are discussed separately below. 

\noindent
\textbf{Case 1.} $s_{k-1}\in \mathcal{B}_{k-1}$:
\begin{align}
    \begin{split}
    &s_{k-1}\in \mathcal{B}_{k-1}\Longrightarrow \mathrm{P}\left( e_3 \right)=0 \Longrightarrow 
    \mathrm{P} \left(e_0^{\mathcal{E}}\right)=0
    \end{split}
\end{align}
and 
\begin{align}
    \begin{split}
        &s_{k-1}\in \mathcal{B}_{k-1}\Longrightarrow \mathbb{O}_{\mathcal{B}_{k-1}}(s_{k-1})=0 \Longrightarrow\\
        &\mathbb{O}_{\mathcal{B}_{k-1}}(s_{k-1}) \leq \mathbb{I}_{\mathcal{A}_{k-1}}(s_{k-1}) \land 
        \mathbb{O}_{\mathcal{B}_{k-1}}(s_{k-1}) \leq\mathbb{E}\left[ V_k^{\mathcal{E}}\left( \mathcal{F}\left(.|s_{k-1},
        \mathcal{E}(s_{k-1},k-1) \right) \right)\right] \Longrightarrow \\
        &\min \left\{\mathbb{O}_{\mathcal{B}_{k-1}}(s_{k-1}), 
        \max\left\{\mathbb{I}_{\mathcal{A}_{k-1}}(s_{k-1}),
        \mathbb{E}\left[ V_k^{\mathcal{E}}\left( \mathcal{F}\left(.|s_{k-1},\mathcal{E}(s_{k-1},k-1) \right) \right)\right]  \right\} \right\}=0
    \end{split}
\end{align}
Therefore, Eq. (\ref{eq8}) holds when $s_{k-1}\in \mathcal{B}_{k-1}$.

\noindent
\textbf{Case 2.} $s_{k-1}\in \mathcal{A}_{k-1}- \mathcal{B}_{k-1}$:
\begin{align}
    \begin{split}
    &s_{k-1}\in \mathcal{A}_{k-1}-\mathcal{B}_{k-1}\Longrightarrow \mathrm{P}\left( e_1 \right)=1 \land \mathrm{P}\left( e_3 \right)=1 \Longrightarrow 
    \mathrm{P} \left(e_0^{\mathcal{E}}\right)=1
    \end{split}
\end{align}
and 
\begin{align}
    \begin{split}
        &s_{k-1}\in \mathcal{A}_{k-1}- \mathcal{B}_{k-1}\Longrightarrow \mathbb{I}_{\mathcal{A}_{k-1}}(s_{k-1})=1 \land 
        \mathbb{O}_{\mathcal{B}_{k-1}}(s_{k-1})=1 \Longrightarrow\\
        &\mathbb{O}_{\mathcal{B}_{k-1}}(s_{k-1}) = \mathbb{I}_{\mathcal{A}_{k-1}}(s_{k-1}) 
         \geq\mathbb{E}\left[ V_k^{\mathcal{E}}\left( \mathcal{F}\left(.|s_{k-1},
        \mathcal{E}(s_{k-1},k-1) \right) \right)\right] \Longrightarrow \\
        &\min \left\{\mathbb{O}_{\mathcal{B}_{k-1}}(s_{k-1}), 
        \max\left\{\mathbb{I}_{\mathcal{A}_{k-1}}(s_{k-1}),
        \mathbb{E}\left[ V_k^{\mathcal{E}}\left( \mathcal{F}\left(.|s_{k-1},\mathcal{E}(s_{k-1},k-1) \right) \right)\right]  \right\} \right\}=1
    \end{split}
\end{align}
Therefore, Eq. (\ref{eq8}) holds when $s_{k-1}\in \mathcal{A}_{k-1}- \mathcal{B}_{k-1}$.

\noindent
\textbf{Case 3.} $s_{k-1}\notin \mathcal{A}_{k-1} \cup \mathcal{B}_{k-1}$:
\begin{align}
    \begin{split}
    &s_{k-1}\notin \mathcal{A}_{k-1} \cup \mathcal{B}_{k-1}\Longrightarrow \mathrm{P}\left( e_1 \right)=0 \land \mathrm{P}\left( e_3 \right)=1 \Longrightarrow 
    \mathrm{P} \left(e_0^{\mathcal{E}}  \right)=\mathrm{P} \left(e_2^{\mathcal{E}}  \right)\\
        &=\mathrm{P}\left[ \left. \left(\exists k_1\in \mathbb{Z}_k^T,\phi_k^T(.|k_1,s_k,\mathcal{E})\in \mathcal{A}_{k_1}\right) \land 
        \left(\forall k_2\in \mathbb{Z}_k^{k_1},\phi_k^{k_1}(.|k_2,s_k,\mathcal{E})\notin \mathcal{B}_{k_2}\right) \right| 
        s_k\sim \mathcal{F}\left(.|s_{k-1},u_{k-1}\right)  \right]\\
        &=\int_{\mathbb{R}^n} V_k^{\mathcal{E}}(s)\mathcal{F}\left(s|s_{k-1},u_{k-1}\right)ds=\mathbb{E}\left[ V_k^{\mathcal{E}}\left( \mathcal{F}\left(.|s_{k-1},\mathcal{E}(s_{k-1},k-1) \right) \right)\right]
    \end{split}
\end{align}
and 
\begin{align}
    \begin{split}
        &s_{k-1}\notin \mathcal{A}_{k-1}\cup \mathcal{B}_{k-1}\Longrightarrow \mathbb{I}_{\mathcal{A}_{k-1}}(s_{k-1})=0 \land 
        \mathbb{O}_{\mathcal{B}_{k-1}}(s_{k-1})=1 \Longrightarrow\\
        &\mathbb{O}_{\mathcal{B}_{k-1}}(s_{k-1}) 
         \geq\mathbb{E}\left[ V_k^{\mathcal{E}}\left( \mathcal{F}\left(.|s_{k-1},
        \mathcal{E}(s_{k-1},k-1) \right) \right)\right] \geq \mathbb{I}_{\mathcal{A}_{k-1}}(s_{k-1}) \Longrightarrow \\
        &\min \left\{\mathbb{O}_{\mathcal{B}_{k-1}}(s_{k-1}), 
        \max\left\{\mathbb{I}_{\mathcal{A}_{k-1}}(s_{k-1}),
        \mathbb{E}\left[ V_k^{\mathcal{E}}\left( \mathcal{F}\left(.|s_{k-1},\mathcal{E}(s_{k-1},k-1) \right) \right)\right]  \right\} \right\}\\
        &=\mathbb{E}\left[ V_k^{\mathcal{E}}\left( \mathcal{F}\left(.|s_{k-1},\mathcal{E}(s_{k-1},k-1) \right) \right)\right]
    \end{split}
\end{align}
Therefore, Eq. (\ref{eq8}) holds when $s_{k-1}\notin \mathcal{A}_{k-1}\cup\mathcal{B}_{k-1}$.

\rightline{$\square$}

Theorem \ref{thm1} reveals a recursive formula, for which the terminal condition can be derived from the following equation:
\begin{align}
    \begin{split}
        V_T^{\mathcal{E}}(s_{T})=  &\mathrm{P} \left[ \left(\exists k_1\in \mathbb{Z}_T^T,\phi_T^T(.|k_1,s_T,\mathcal{E})\in \mathcal{A}_{k_1}\right) \land 
        \left(\forall k_2\in \mathbb{Z}_T^{k_1},\phi_T^{k_1}(.|k_2,s_k,\mathcal{E})\notin \mathcal{B}_{k_2}\right)  \right]\\
        =&\mathrm{P} \left( s_T\in \mathcal{A}_T \land s_T\notin \mathcal{B}_T  \right)=
        \min \left[ \mathbb{I}_{\mathcal{A}_T}\left( s_T \right),\mathbb{O}_{\mathcal{B}_T}\left( s_T \right)  \right]
    \end{split}
\end{align}
Consequently, the complete backward recursive procedure is as follows:
\begin{align}
    \begin{split}
        &V_T^{\mathcal{E}}(s_{T})=\min \left[ \mathbb{I}_{\mathcal{A}_T}\left( s_T \right),\mathbb{O}_{\mathcal{B}_T}\left( s_T \right)  \right]\\
        &V_{k-1}^{\mathcal{E}}(s_{k-1})=\min \left\{\mathbb{O}_{\mathcal{B}_{k-1}}(s_{k-1}), 
        \max\left\{\mathbb{I}_{\mathcal{A}_{k-1}}(s_{k-1}),
        \mathbb{E}\left[ V_k^{\mathcal{E}}\left( \mathcal{F}\left(.|s_{k-1},\mathcal{E}(s_{k-1},k-1) \right) \right)\right]  \right\} \right\}\\
        &...\\
        &V_{0}^{\mathcal{E}}(s_0)=\left\{\mathbb{O}_{\mathcal{B}_{0}}(s_{0}), 
        \max\left\{\mathbb{I}_{\mathcal{A}_{0}}(s_{0}),
        \mathbb{E}\left[ V_1^{\mathcal{E}}\left( \mathcal{F}\left(.|s_{0},\mathcal{E}(s_{0},0) \right) \right)\right]  \right\} \right\}
    \end{split}
\end{align}
According to Theorem \ref{thm1}, $V_{0}^{\mathcal{E}}(s_0)=\mathrm{P}\left(s_0|T,\mathcal{E}\right)$, and the 
$\gamma-$probabilistic reachable set can be characterized by the $\gamma-$level set of $V_{0}^{\mathcal{E}}(.)$, i.e.:
\begin{align}
    \mathcal{R}_\gamma^\mathcal{E}\left(\mathcal{A},\mathcal{B},T\right)=
    \left\{s_0\in\mathbb{R}^n|V_{0}^{\mathcal{E}}(s_0)\geq\gamma \right\}
\end{align}

\begin{thm}\label{thm2}
    Let the scalar function $V_k^{\mathrm{max}}(s_k)$
    denote the probability that, under the optimal control policy, the evolution of the system from state $s_k$ at time $k$ 
    reaches the target set at some time in $\mathbb{Z}_k^T$ and avoids the obstacle, i.e.:
    \begin{align}
        \begin{split}
        V_k^{\mathrm{max}}(s_{k})=\max_{\mathcal{E}}  \mathrm{P} \left[ \left(\exists k_1\in \mathbb{Z}_k^T,\phi_k^T(.|k_1,s_k,\mathcal{E})\in \mathcal{A}_{k_1}\right) \land 
        \left(\forall k_2\in \mathbb{Z}_k^{k_1},\phi_k^{k_1}(.|k_2,s_k,\mathcal{E})\notin \mathcal{B}_{k_2}\right)  \right]
        \end{split}
    \end{align}
    If 
    \begin{align}
        \begin{split}
            V_{k-1}^{\mathrm{max}}(s_{k-1})=\min \left\{\mathbb{O}_{\mathcal{B}_{k-1}}(s_{k-1}), 
            \max\left\{\mathbb{I}_{\mathcal{A}_{k-1}}(s_{k-1}),
            \max_{u_{k-1}\in\mathcal{U}} \mathbb{E}\left[ V_k^{\mathrm{max}}\left( \mathcal{F}\left(.|s_{k-1},u_{k-1} \right) \right)\right]  \right\} \right\}
        \end{split}
    \end{align}
    Then 
    \begin{align}\label{eq25}
        \begin{split}
        V_{k-1}^{\mathrm{max}}(s_{k-1})= \max_{\mathcal{E}}  \mathrm{P} \left[ \left(\exists k_1\in \mathbb{Z}_{k-1}^T,
        \phi_{k-1}^T(.|k_1,s_{k-1},\mathcal{E})\in \mathcal{A}_{k_1}\right) \right. \land \\
        \left.\left(\forall k_2\in \mathbb{Z}_{k-1}^{k_1},\phi_{k-1}^{k_1}(.|k_2,s_{k-1},\mathcal{E})\notin \mathcal{B}_{k_2}\right)  \right]
        \end{split}
    \end{align}
\end{thm}

\noindent
\textbf{Proof.} The proof of Theorem \ref{thm2} is similar to that of Theorem \ref{thm1}.
Here, we continue with the notations in Eq. (\ref{eq1}).
Eq. (\ref{eq25}) can be expanded as follows:
\begin{align}
    V_{k-1}^{\mathrm{max}}(s_{k-1})=\max_{\mathcal{E}} \mathrm{P}\left(e_0^{\mathcal{E}}\right)
    =\left\{ 1- \left[ 1-\mathrm{P}\left( e_1 \right) \right] \times  
    \left[1- \max_{\mathcal{E}} \mathrm{P}\left( e_2^{\mathcal{E}}  \right) \right] \right\}\times 
    \mathrm{P}\left( e_3 \right)
\end{align}
Case 1 ($s_{k-1}\in \mathcal{B}_{k-1}$) and Case 2 
($s_{k-1}\in\mathcal{A}_{k-1}- \mathcal{B}_{k-1}$) are exactly the same as those in Theorem \ref{thm1}, 
and will not be repeated here. Case 3 is discussed below.

\noindent
\textbf{Case 3.} $s_{k-1}\notin \mathcal{A}_{k-1} \cup \mathcal{B}_{k-1}$:
\begin{align}
    \begin{split}
    &s_{k-1}\notin \mathcal{A}_{k-1} \cup \mathcal{B}_{k-1}\Longrightarrow \mathrm{P}\left( e_1 \right)=0 \land \mathrm{P}\left( e_3 \right)=1 \Longrightarrow 
    \max_{\mathcal{E}}\mathrm{P} \left(e_0^{\mathcal{E}}  \right)=\max_{\mathcal{E}}\mathrm{P} \left(e_2^{\mathcal{E}}  \right)\\
    &=\max_{\mathcal{E},u_{k-1}}\mathrm{P}\left[ \left. \left(\exists k_1\in \mathbb{Z}_k^T,\phi_k^T(.|k_1,s_k,\mathcal{E})\in \mathcal{A}_{k_1}\right) \land 
    \left(\forall k_2\in \mathbb{Z}_k^{k_1},\phi_k^{k_1}(.|k_2,s_k,\mathcal{E})\notin \mathcal{B}_{k_2}\right) \right| 
    s_k\sim \mathcal{F}\left(.|s_{k-1},u_{k-1}\right)  \right]\\
    &=\max_{u_{k-1}}\int_{\mathbb{R}^n} V_k^{\mathrm{max}}(s)\mathcal{F}\left(s|s_{k-1},u_{k-1}\right)ds=
    \max_{u_{k-1}}\mathbb{E}\left[ V_k^{\mathrm{max}}\left( \mathcal{F}\left(.|s_{k-1},u_{k-1} \right) \right)\right]
    \end{split}
\end{align}
and 
\begin{align}
    \begin{split}
        &s_{k-1}\notin \mathcal{A}_{k-1}\cup \mathcal{B}_{k-1}\Longrightarrow \mathbb{I}_{\mathcal{A}_{k-1}}(s_{k-1})=0 \land 
        \mathbb{O}_{\mathcal{B}_{k-1}}(s_{k-1})=1 \Longrightarrow\\
        &\mathbb{O}_{\mathcal{B}_{k-1}}(s_{k-1}) 
         \geq\mathbb{E}\left[ V_k^{\mathcal{E}}\left( \mathcal{F}\left(.|s_{k-1},
        \mathcal{E}(s_{k-1},k-1) \right) \right)\right] \geq \mathbb{I}_{\mathcal{A}_{k-1}}(s_{k-1}) \Longrightarrow \\
        &\min \left\{\mathbb{O}_{\mathcal{B}_{k-1}}(s_{k-1}), 
        \max\left\{\mathbb{I}_{\mathcal{A}_{k-1}}(s_{k-1}),
        \max_{u_{k-1}}\mathbb{E}\left[ V_k^{\mathrm{max}}\left( \mathcal{F}\left(.|s_{k-1},u_{k-1} \right) \right)\right]  \right\} \right\}\\
        &=\max_{u_{k-1}}\mathbb{E}\left[ V_k^{\mathrm{max}}\left( \mathcal{F}\left(.|s_{k-1},u_{k-1} \right) \right)\right]
    \end{split}
\end{align}
Therefore, Eq. (\ref{eq25}) holds when $s_{k-1}\notin \mathcal{A}_{k-1}\cup\mathcal{B}_{k-1}$.

\rightline{$\square$}

Similarly, Theorem \ref{thm2} indicates that, the maximum $\gamma-$probabilistic reachable set 
can be represented by the $\gamma-$level set of $V_0^{\mathrm{max}}(.)$, i.e.:
\begin{align}
    \mathcal{R}_\gamma^{\mathrm{max}}(\mathcal{A},\mathcal{B},T)=\left\{s_0\in\mathbb{R}^n|V_{0}^{\mathrm{max}}(s_0)\geq\gamma \right\}
\end{align}
and $V_0^{\mathrm{max}}(.)$ can be obtained by the following recursive formula:
\begin{align}\label{eq30}
    \begin{split}
        &V_T^{\mathrm{max}}(s_{T})=\min \left[ \mathbb{I}_{\mathcal{A}_T}\left( s_T \right),\mathbb{O}_{\mathcal{B}_T}\left( s_T \right)  \right]\\
        &V_{k-1}^{\mathrm{max}}(s_{k-1})=\min \left\{\mathbb{O}_{\mathcal{B}_{k-1}}(s_{k-1}), 
        \max\left\{\mathbb{I}_{\mathcal{A}_{k-1}}(s_{k-1}),
        \max_{u_{k-1}\in\mathcal{U}} \mathbb{E}\left[ V_k^{\mathrm{max}}\left( \mathcal{F}\left(.|s_{k-1},u_{k-1} \right) \right)\right]  \right\} \right\}\\
        &...\\
        &V_{0}^{\mathrm{max}}(s_0)=\left\{\mathbb{O}_{\mathcal{B}_{0}}(s_{0}), 
        \max\left\{\mathbb{I}_{\mathcal{A}_{0}}(s_{0}),
        \max_{u_{0}\in\mathcal{U}}\mathbb{E}\left[ V_1^{\mathrm{max}}\left( \mathcal{F}\left(.|s_{0},u_0 \right) \right)\right]  \right\} \right\}
    \end{split}
\end{align}
In addition, Eq. (\ref{eq30}) reveals the expression for the optimal control policy:
\begin{align}\label{eq31}
    \mathcal{E}^*\left(s_k,k\right)=\arg\max_{u_k \in \mathcal{U}} \mathbb{E} \left[ V_{k+1}^{\mathrm{max}}\left( \mathcal{F}\left(.|s_{k},u_k \right) \right)\right]
\end{align}

\section{Method implementation}
For any $k\in \mathbb{Z}_0^{T-1}$, the analytic forms of the expressions of 
$V_k^{\mathcal{E}}(.)$ and $V_k^{\mathrm{max}}(.)$ are difficult to obtain. 
This section introduces a method based on grid interpolation
to approximate these functions. First, a rectangular
computational domain, denoted as $\mathcal{S}$, needs to be specified in
the state space. Then, divide $\mathcal{S}$ into a Cartesian grid structure.
The values of function $V_k^{\mathcal{E}}(.)$ or $V_k^{\mathrm{max}}(.)$ at the grid points are stored in
an array with the same dimensions as the state space, and $V_k^{\mathcal{E}}(.)$ or $V_k^{\mathrm{max}}(.)$ 
can be approximated by the grid interpolation. 

In addition, for any $k\in \mathbb{Z}_0^{T-1}$, 
both of the recursive procedures described above involve computing the expectation value 
$\mathbb{E}\left[ V_k^{\mathcal{E}}\left( \mathcal{F}\left(.|s_{k-1},\mathcal{E}(s_{k-1},k-1) \right) \right)\right]$ or 
$\mathbb{E}\left[ V_k^{\mathrm{max}}\left( \mathcal{F}\left(.|s_{k-1},u_{k-1} \right) \right)\right]$. 
In the current study, these expectation values will be calculated by using Monte Carlo method.
Denote the sampling result from probability distribution $\mathcal{F}\left(.|s_{k-1},\mathcal{E}(s_{k-1},k-1) \right)$ or 
$\mathcal{F}\left(.|s_{k-1},u_{k-1} \right)$ as $\{s^1,...,s^m\}$, then the expectation values
can be estimated by the following equations:
\begin{align}
    \begin{split}
        &\mathbb{E}\left[ V_k^{\mathcal{E}}\left( \mathcal{F}\left(.|s_{k-1},
        \mathcal{E}(s_{k-1},k-1) \right) \right)\right]=\frac{1}{m}\sum_{i=1}^m V_k^{\mathcal{E}}\left(s^i\right)  \\
        &\mathbb{E}\left[ V_k^{\mathrm{max}}\left( \mathcal{F}\left(.|s_{k-1},u_{k-1} \right) \right)\right]
        =\frac{1}{m}\sum_{i=1}^m V_k^{\mathrm{max}}\left(s^i\right)
    \end{split}
\end{align}

Given all the techniques described above, the complete algorithm can be obtained. 
As an example, take the maximum probabilistic reachable set of a two-dimensional system, 
where the system state is denoted as $s=[x,y]^\mathrm{T}$, 
the pseudocode of the proposed method
is shown in Algorithm \ref{alg1}.

\begin{algorithm}[htbp]
	\label{alg1}
    \caption{Method to compute maximum probabilistic reachable set}
	\begin{algorithmic}[1]
	  \State \textbf{Inputs:}Time horizon $T$, dynamic system $s_{k+1}\sim \mathcal{F}(.|s_k,u_k)$, 
        a real number $\gamma\in [0,1]$, computational domain $\mathcal{S}=[x_d,x_u]\times [y_d,y_u]$, grids number $N_x\times N_y$;
	  \State Construct two $N_x\times N_y$ arrays, denoted as $\mathcal{V}$ and $\mathcal{V}'$;
	  \State $\displaystyle{\delta x\leftarrow \frac{ x_u-x_d }{ N_x-1 },\delta y\leftarrow \frac{ y_u-y_d }{ N_y-1 }}$;
      \For{$i_x\leftarrow 0,...,N_x-1 $} \textcolor{gry}{\quad\quad\quad\quad\quad $\backslash\backslash$ The terminal condition of the recursive formula.}
        \For{$i_y\leftarrow 0,...,N_y-1 $}
            \State $s_T\leftarrow \left[x_d+i_x\delta x,y_d+i_y\delta y    \right]^\mathrm{T}$;
            \State $\mathcal{V}[i_x][i_y]\leftarrow \min\left[ \mathbb{O}_{\mathcal{B}_T}\left(s_T\right),\mathbb{I}_{\mathcal{A}_T}\left(s_T\right)   \right]$
        \EndFor
      \EndFor

	  \For{$k \leftarrow N-1,...,0$}
      \For{$i_x\leftarrow 0,...,M_x-1 $}
        \For{$i_y\leftarrow 0,...,M_y-1 $}
          \State $s_k\leftarrow \left[x_d+i_x\delta x,y_d+i_y\delta y    \right]^\mathrm{T}$;
          \If{$s_k \in \mathcal{B}_k$}\textcolor{gry}{\quad\quad\quad\quad\quad $\backslash\backslash$ $s_k \in \mathcal{B}_k \Longrightarrow V_k(s_k)=\mathbb{O}_{\mathcal{B}_k}(s_k)=0$}
            \State $\mathcal{V}'[i_x][i_y]\leftarrow 0$;
            \State Continue;
          \EndIf
          \If{$s_k \in \mathcal{A}_k$}\textcolor{gry}{\quad\quad\quad\quad\quad $\backslash\backslash$ $s_k \in \mathcal{A}_k-\mathcal{B}_k \Longrightarrow V_k(s_k)=\mathbb{I}_{\mathcal{A}_k}(s_k)=1$}
            \State $\mathcal{V}'[i_x][i_y]\leftarrow 1$;
            \State Continue;
          \EndIf
          \State Construct a bilinear interpolation function $\widehat{V}(.)$ using $\mathcal{V}$;
          \State $v^*\leftarrow 0$;
          \For{$u_k\in \mathcal{U}$}
            \State Sample from probability distribution $\mathcal{F}(.|s_k,u_k)$, denote the samples as $\{s^1,...,s^m\}$;
            \State $v\leftarrow \frac{1}{m} \sum_{i=1}^m \widehat{V}(s^i)$;
            \If{$v^*<v$}
              \State $v^*\leftarrow v$;
            \EndIf
          \EndFor
          \State $\mathcal{V}'[i_x][i_y]\leftarrow v^*$;
          \textcolor{gry}{\quad\quad\quad\quad\quad $\backslash\backslash$ 
          $s_k \notin \mathcal{A}_k\cup \mathcal{B}_k \Longrightarrow V_k^{\mathrm{max}}(s_k)=\displaystyle{\max_{u_k}} \mathbb{E} \left[ V_{k+1}^{\mathrm{max}}( \mathcal{F}(.|s_k,u_k) ) \right]$}
        \EndFor
      \EndFor
      \State Copy $\mathcal{V}'$ to $\mathcal{V}$;
    \EndFor
    \State Return $\left\{s\in\Omega|\widehat{V}(s)\geq \gamma \right\}$;
    \end{algorithmic}
\end{algorithm}

\section{Numerical examples}
A flight vehicle moves in a plane, the vehicle
is modeled as a simple mass point with fixed linear velocity $v=1$ and controllable heading angular velocity.
Its motion in still air is determined by the following equation:
\begin{align}
	\left\{\begin{array}{l}
		\dot{x}=v \cos \theta \\
		\dot{y}=v \sin \theta \\
		\dot{\theta}=u
	\end{array}\right.
\end{align}
where $[x,y]^\mathrm{T} \in \mathbb{R}^2$ and $\theta\in (-\pi,\pi]$ are the position and heading angle of the vehicle respectively,
and $u\in \mathcal{U}=[-1,1]$ is the control input.
There exists a wind field in the plane, which is determined by the following vector field:
\begin{align}
	\left[
		\begin{array}{c}
			w_x(x,y)\\w_y(x,y)
		\end{array}
	\right]=\left[
		\begin{array}{c}
			\displaystyle{-y-0.1y^3} \\ \displaystyle{x+0.1x^3}
		\end{array}
	\right]
\end{align}
Then the dynamics of the vehicle in the wind field can be described by:
\begin{align}\label{eq35}
	\dot{s}=f(s,u)=\left[
		\begin{array}{c}
			v \cos \theta +w_x(x,y)\\ v\sin \theta+w_y(x,y) \\ u
		\end{array}
	\right]
\end{align}
The vehicle selects a control input from $\mathcal{U}$ every $\Delta t=0.1$ time,
and the control input is fixed as a constant during each time step.
The uncertainty of the vehicle's position accumulated in one time step is represented by a uniform distribution
over a circular region with radius $r=0.1$. Therefore, the time discrete form of the system is as follows:
\begin{align}
    s_{k+1}\sim \mathcal{F}\left(.|s_k,u_k\right)=F(s_k,u_k)+\mathscr{U}_r(.)
\end{align}
where $F(s_k,u_k)$ is derived by solving Eq. (\ref{eq35}) using Runge-Kutta method,
$\mathscr{U}_r(.)$ is a probability distribution whose probability density at $s=[x,y,\theta]^\mathrm{T}$ is:
\begin{align}
    \mathscr{U}_r(s)=\begin{cases}
        \displaystyle{ \frac{1}{\pi r^2}},& x^2+y^2\leq r^2\\
        0,& x^2+y^2> r^2
    \end{cases}
\end{align}

\begin{figure}[htbp]
    \centering
    \includegraphics[width=0.5\textwidth]{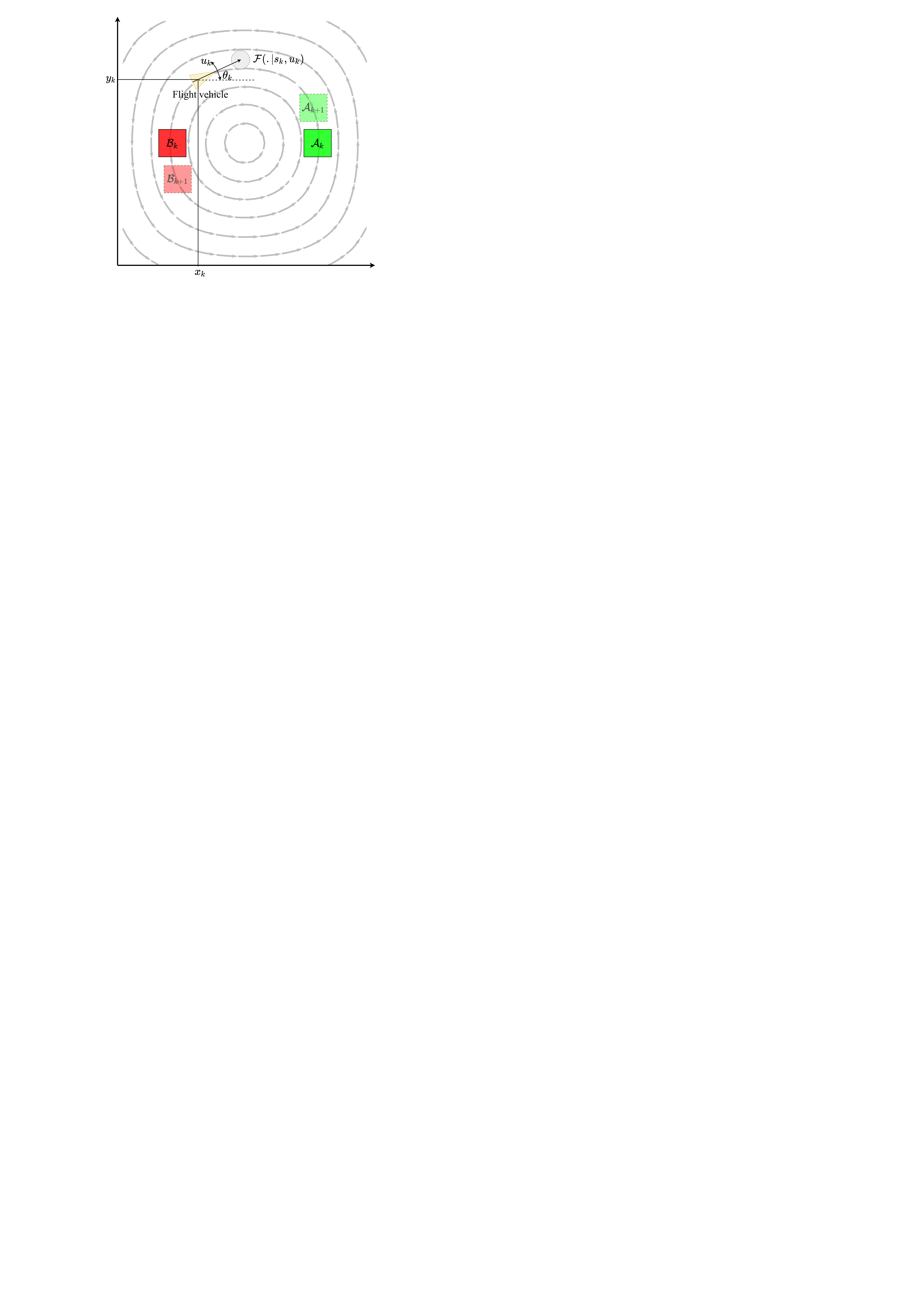}
    \caption{Sketch of the numerical example.}
    \label{fig1}
\end{figure}

The target set and obstacle at the $k$th time step are:
\begin{align}
    \begin{split}
      &\mathcal{A}_k=\left\{[x,y,\theta]^\mathrm{T} \bigg|
      x\in \left[ 2 \cos\left( \frac{k\pi}{40}  \right)-\frac{1}{2},2 \cos\left( \frac{k\pi}{40}  \right)+\frac{1}{2}  \right],
        y\in \left[ 2 \sin\left( \frac{k\pi}{40}  \right)-\frac{1}{2},2 \sin\left( \frac{k\pi}{40}  \right)+\frac{1}{2}  \right] \right\}\\
      &\mathcal{B}_k=\left\{[x,y,\theta]^\mathrm{T} \bigg|
      x\in \left[ 2 \cos\left( \frac{k\pi}{40}+\pi  \right)-\frac{1}{2},2 \cos\left( \frac{k\pi}{40} +\pi  \right)+\frac{1}{2}  \right],\right.\\
      &\quad \quad \quad \quad \quad \quad \quad \left. y\in \left[ 2 \sin\left( \frac{k\pi}{40}+\pi \right)-\frac{1}{2},2 \sin\left( \frac{k\pi}{40}+\pi   \right)+\frac{1}{2}  \right] \right\}
    \end{split}
\end{align}
and the time horizon is $T=23$, the given probability is $\gamma=0.6$.
The problem is depicted visually in Fig. \ref{fig1}.

\subsection{Computation of probabilistic reachable set}
Assume that the control policy aims to point the vehicle's heading to the center of the target set. 
That is, the control policy is fixed at:
% \begin{align}
%     \mathcal{E}\left(s_k,u_k\right)=\begin{cases}
%         1, & \displaystyle{ \text{atan2}\left[\sin\left( \frac{k\pi}{40}  \right)-y_k, 2 \cos\left( \frac{k\pi}{40}  \right)-x_k \right]-\theta_k>0 }\\
%         -1, & \displaystyle{ \text{atan2}\left[\sin\left( \frac{k\pi}{40}  \right)-y_k, 2 \cos\left( \frac{k\pi}{40}  \right)-x_k \right]-\theta_k<0 }\\
%         0,& \displaystyle{ \text{atan2}\left[\sin\left( \frac{k\pi}{40}  \right)-y_k, 2 \cos\left( \frac{k\pi}{40}  \right)-x_k \right]-\theta_k=0 }
%     \end{cases}
% \end{align}
\begin{align}\label{eq39}
    \mathcal{E}\left(s_k,k\right)=\min\left\{1,\max\left[-1, \displaystyle{ \text{atan2}\left[\sin\left( \frac{k\pi}{40}  \right)-y_k, 2 \cos\left( \frac{k\pi}{40}  \right)-x_k \right]-\theta_k }  \right]\right\}
\end{align}

where the definition of function $\text{atan2}$ is \cite{a20}:
\begin{align}
    \begin{split}
      \text{atan2}(\Delta y,\Delta x)=
      \begin{cases}
          \arctan \frac{\Delta y}{\Delta x}, &\Delta x>0\\
          \arctan \frac{\Delta y}{\Delta x}+\pi, &\Delta x<0 \text{ and } \Delta y\geq 0\\
          \arctan \frac{\Delta y}{\Delta x}-\pi, &\Delta x<0 \text{ and } \Delta y< 0\\
          \frac{\pi}{2}, &\Delta x=0 \text{ and } \Delta y>0\\
          -\frac{\pi}{2}, &\Delta x=0 \text{ and } \Delta y<0
      \end{cases}
    \end{split}
\end{align}

\begin{figure}[htbp]
    \centering
    \subfigure[$\gamma-$probabilistic reachable set.]{\includegraphics[width=0.4\textwidth]{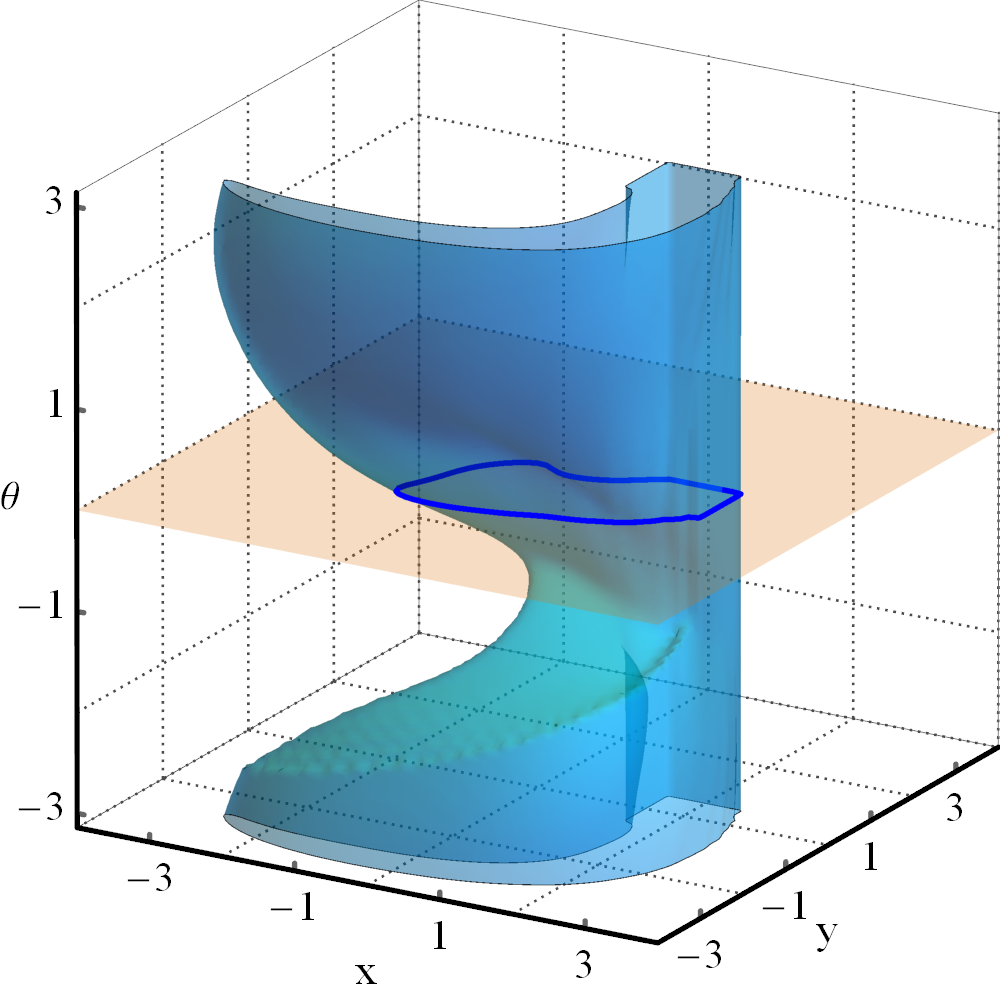}}\ \ \ \ \ \ \ \ \ \  \ \ \ \ \
    \subfigure[Simulation results in slice $\theta=0$.]{\includegraphics[width=0.4\textwidth]{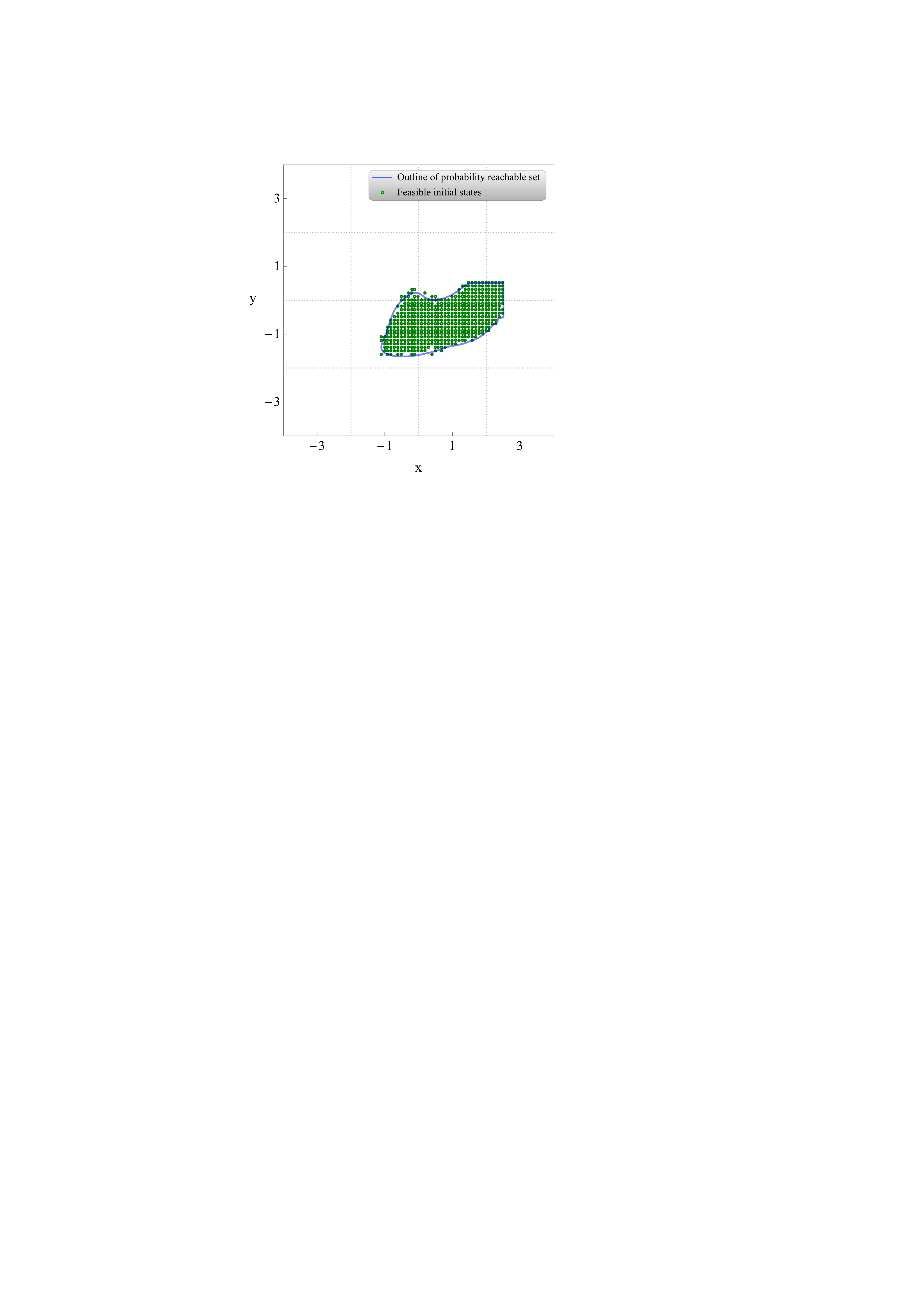}}
    \caption{Computation results of $\gamma-$probabilistic reachable set.}
    \label{fig2}
\end{figure}

The solver setups of this example are listed in Table \ref{tb1}.
\begin{table}[htbp]
	\centering
	\caption{Solver settings for the computation of probabilistic reachable set}
	\label{tb1}
	\begin{tabular}{lc}
	\toprule[1pt] 
	\textbf{Parameter} 			  			& \textbf{Setting}  \\ \toprule
	Computational domain $\mathcal{S}$         	&      $[-4,4]\times [-4,4] \times [-\pi,\pi]$      \\   
	Number of grid points $N_x\times N_y \times N_\theta$   &      $201\times 201 \times 201$     \\  
	Time horizon $T$              				&      $23$      \\    
    Probability $\gamma$              				&      $0.6$      \\   
	Sample size for the Monte Carlo method $m$            	&      10000     \\ \bottomrule[1pt] 
	\end{tabular}
\end{table}
The computation result of the $\gamma-$probabilistic reachable set is shown in Fig. \ref{fig2}(a).
In order to verify the correctness of the result, some points in slice $\theta=0$ 
of the state space are chosen as the initial states, for each initial state, 1000 times of evolution under control policy 
\ref{eq39} are simulated and the counts of avoiding obstacles and reaching the target set are recorded.
Fig. \ref{fig2}(b) shows the simulation results,
where the evolutions starting from the points marked by the green circles reach the target set and avoid the obstacles more than 600 times.
It can be seen that the outline of $\gamma-$probabilistic
reachable sets and the border of the area marked
by the green circles almost coincide, which indicates the accuracy of the computation results. 

\subsection{Computation of Maximum probabilistic reachable set}
\begin{figure}[htbp]
    \centering
    \subfigure[Maximum $\gamma-$probabilistic reachable set.]{\includegraphics[width=0.4\textwidth]{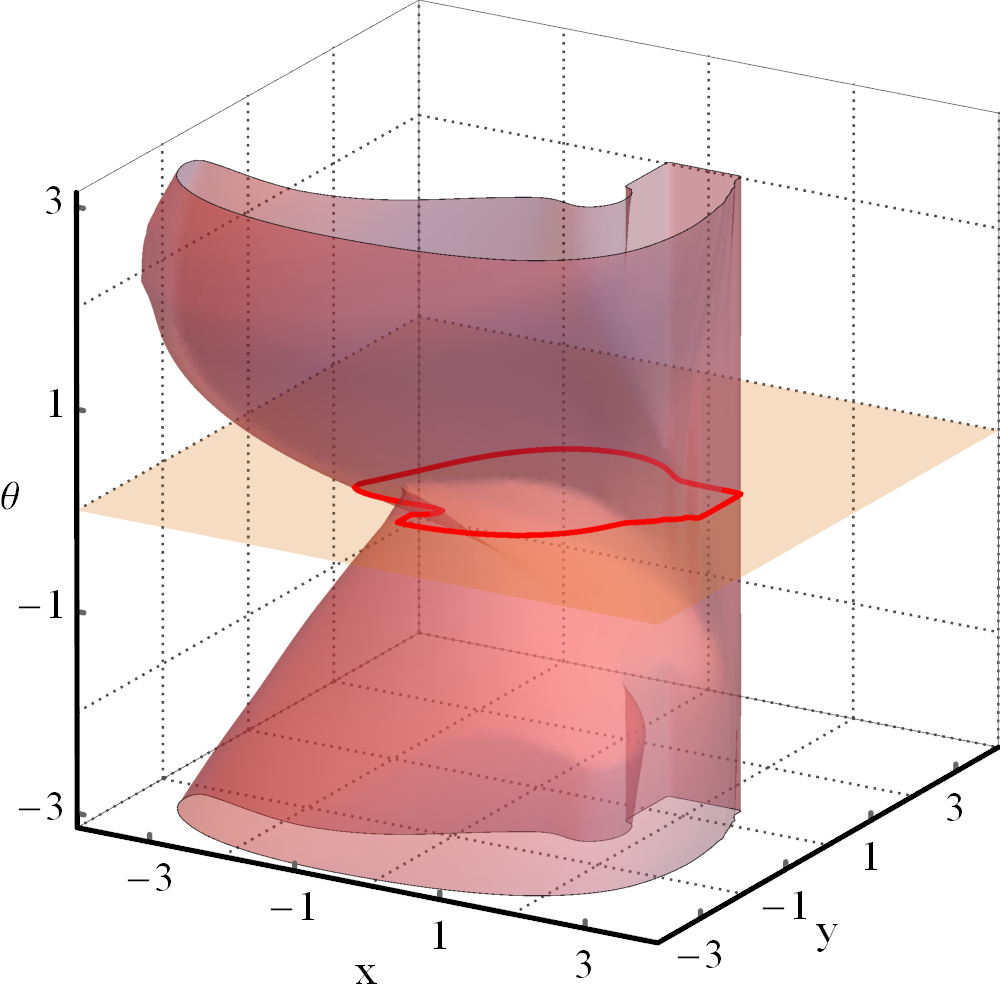}}\ \ \ \ \ \ \ \ \ \  \ \ \ \ \
    \subfigure[Simulation results in slice $\theta=0$.]{\includegraphics[width=0.4\textwidth]{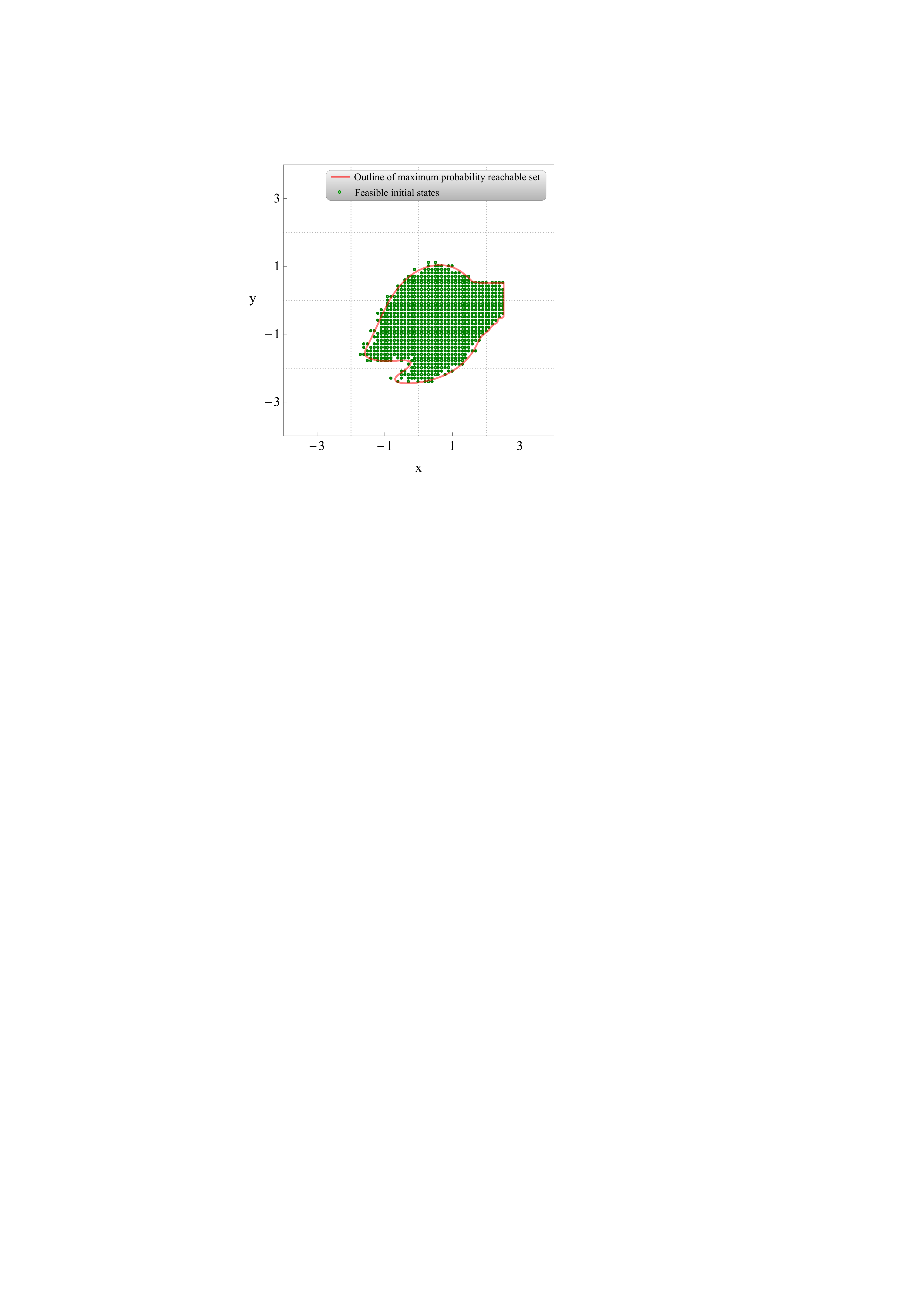}}
    \caption{Computation results of maximum $\gamma-$probabilistic reachable set.}
    \label{fig3}
\end{figure}

From line 24 of Algorithm \ref{alg1}, 
it is clear that the computation of the maximum probability reachable set requires traversing the set $\mathcal{U}$.
In this example, since the set $\mathcal{U}$ is a continuous set, it is impractical to traverse this set.
Therefore, the $\mathcal{U}$ is replaced by the following set consisting of some discrete points in the computation procedure.
\begin{align}
    \bar{\mathcal{U}}=\left\{-1+\frac{1}{10}i \bigg| i=0,1,...,20 \right\}
\end{align}
The solver setups are also shown in Table \ref{tb1}.
The computation result of the maximum $\gamma-$probabilistic reachable set is shown in Fig. \ref{fig3}(a). 
During the computation, the functions $V^{\mathrm{max}}_k(.)$ are saved for any $k\in \mathbb{Z}_0^T$. 
These functions can be used to implement the control policy (\ref{eq31}).
In this subsection, the initial states inside slice $\theta=0$ are also simulated to validate the correctness of the result.
In these simulations, the control policy is chosen as Eq. (\ref{eq31}).
Fig. \ref{fig3}(b) displays the simulation results, which also illustrate the accuracy of the proposed method.

\section{Conclusions}
In this paper, we study the reachability problems of non-deterministic discrete-time systems. 
Based on the existing works, the time-dependent target set and obstacle are taken into account and
the definition of probabilistic reachable set is refined.
Two types of probabilistic reachable sets are discussed, namely, probabilistic
reachable set with a given control policy and maximum probabilistic reachable set.
To compute these probabilistic reachable sets, a numerical method is proposed.
In the proposed method, the probability reachable set is represented as 
a non-zero level set of a scalar function that is approximated by backward recursion and grid interpolation.
The scalar function generated by each recursive step can also be used to design optimal control policy.
The paper concludes with some examples to verify the effectiveness of the proposed method.

The proposed method has some potential for
improvement. For example, the computation of expectation values using Monte Carlo method 
is computationally expensive
The sample size required grows as the accuracy requirement increases.
New mechanisms for computing expectations will be developed in our future work.

% Then the following theorem can be obtained:
% \begin{thm}
% 	For any $s_0\in \mathcal{R}_c(K,J)$, and $J< \lambda \bar{T}+\Lambda$, 
% 	the trajectory initialized from $s_0$ can reach the target set $K$ under control law (\ref{cl2}) before the 
% 	performance index increasing to $J$.
% \end{thm}

% \begin{prf}
% 	For a small enough time step size $\Delta t$, the optimal control input should satisfy the following equation:
% 	\begin{align}
% 		u^*(s_0)=\arg\min_{u\in\mathcal{U}} \left[ \widetilde{W}_c(s_0+f(s_0,u)\Delta t)+ c(s_0,u)\Delta t    \right]
% 	\end{align}
% 	Denote $s_0'=s_0+f(s_0,u^*)\Delta t$, then 
% 	\begin{align}
% 		\label{thm3eq2}
% 		\begin{split}
% 			&\widetilde{W}_c(s_0')+c(s_0,u^*)=\widetilde{W}_c(s_0)\Delta t \\
% 			&\Longrightarrow \widetilde{W}_c(s_0') < \widetilde{W}_c(s_0)
% 		\end{split}
% 	\end{align}
% 	According to Theorem \ref{thm3}, we have: 
% 	\begin{align}
% 		\begin{split}
% 			&s_0\in \mathcal{R}_c(K,J) \land J < \lambda \bar{T}+\Lambda  \\
% 			&\Longrightarrow \widetilde{W}_c(s_0)\leq J< \lambda \bar{T}+\Lambda 
% 		\end{split}
% 	\end{align}
% 	Therefore,
% 	\begin{align}
% 		\begin{split}
% 			&\widetilde{W}_c(s_0')< J< \lambda \bar{T}+\Lambda \Longrightarrow  \\
% 			&s_0'\in \left\{s\in\mathbb{R}^n| \widetilde{W}_c(s)< \lambda \bar{T}+\Lambda\right\}
% 		\end{split}
% 	\end{align}

% \end{prf}

\section*{Acknowledgements}
\par The authors gratefully acknowledge support from National Defense Outstanding Youth Science Foundation (Grant No. 2018-JCJQ-ZQ-053), 
and Central University Basic Scientific Research Operating Expenses Special Fund Project Support (Grant No. NF2018001).
Also, the authors would like to thank
the anonymous reviewers, associate editor, and editor for
their valuable and constructive comments and suggestions.

\bibliographystyle{plain}
% \bibliography{Bibliography/IEEEabrv,Bibliography/BIB_xx-TIE-xxxx}\ %IEEEabrv instead of IEEEfull
\bibliography{mybib}

\end{document}